\documentclass[twocolumn,showpacs,aps,prl,superscriptaddress,letterpaper]{revtex4}
\usepackage{graphicx}
\usepackage{dcolumn}
\usepackage{amsmath}
\usepackage{epsfig}
\usepackage{multirow}
\usepackage{subfigure}

\long\def\inst#1{\par\nobreak\kern 4pt\nobreak
    {\itshape #1}\par\vskip 10pt plus 3pt minus 3pt}
\RequirePackage{xspace}
\usepackage{relsize}

\newcommand{\BABARPubYear}    {09}
\newcommand{\BABARPubNumber}  {015}

\newcommand{\SLACPubNumber} {13653}

\input babarsym
\def\cpoddhiggs      {\ensuremath{{A^{0}}}\xspace}

\def\n1Spipi     {\ensuremath{}\xspace}

\def\beq{\begin{equation}}
\def\eeq{\end{equation}}
\def\bea{\begin{eqnarray}}
\def\eea{\end{eqnarray}}
\def\bq{\begin{quote}}
\def\eq{\end{quote}}
\def\bi{\begin{itemize}}
\def\ei{\end{itemize}}
\def\bc{\begin{center}}
\def\ec{\end{center}}

\newcommand{\eg}{{\em e.g.}}

\def\etal{{\em et al.}}

\newcommand{\higgsmass}{\ensuremath{m_{A^0}}\xspace}

\newcommand{\redmass}{\ensuremath{m_R}\xspace}

\begin{document}

\onecolumngrid
\begin{flushleft}
\babar-PUB-\BABARPubYear/\BABARPubNumber \\
SLAC-PUB-\SLACPubNumber \\
%arXiv:\LANLNumber [hep-ex] \\
% May 27, 2009 \\
\end{flushleft}

\vspace{-\baselineskip}
%%\twocolumngrid
% Title of the paper
\title{
\large \bfseries \boldmath
Search for Dimuon Decays of a Light Scalar Boson in Radiative
Transitions $\Upsilon\to\gamma\cpoddhiggs$
}

%%%%%%%%%%%%%%%%%%%%%%%%%%%%%%%%%%%%%%%%%%%%%%%%%%%%%%%%%%%%%%%%%%%%%%%%%%%%%%%%%%
%% author list :
%
%% author list as of 03-Apr-2009 (488 authors)
%
\author{B.~Aubert}
\author{Y.~Karyotakis}
\author{J.~P.~Lees}
\author{V.~Poireau}
\author{E.~Prencipe}
\author{X.~Prudent}
\author{V.~Tisserand}
\affiliation{Laboratoire d'Annecy-le-Vieux de Physique des Particules (LAPP), Universit\'e de Savoie, CNRS/IN2P3,  F-74941 Annecy-Le-Vieux, France}
\author{J.~Garra~Tico}
\author{E.~Grauges}
\affiliation{Universitat de Barcelona, Facultat de Fisica, Departament ECM, E-08028 Barcelona, Spain }
\author{M.~Martinelli$^{ab}$}
\author{A.~Palano$^{ab}$ }
\author{M.~Pappagallo$^{ab}$ }
\affiliation{INFN Sezione di Bari$^{a}$; Dipartimento di Fisica, Universit\`a di Bari$^{b}$, I-70126 Bari, Italy }
% ...
\collaboration{The \babar\ Collaboration}
\noaffiliation

\date{May 27, 2009}

%%%%%%%%%%%%%%%%%%%%%%%%%%%%%%%%%%%%%%%%%%%%%%%%%%%%%%%%%%%%%%%%%%%%%%%%%%%%%%%%%%
%                             Abstract                                          %%
%%%%%%%%%%%%%%%%%%%%%%%%%%%%%%%%%%%%%%%%%%%%%%%%%%%%%%%%%%%%%%%%%%%%%%%%%%%%%%%%%%

\begin{abstract}
We search for evidence of a light scalar boson in the
radiative decays of 
the \Y2S and \Y3S resonances: $\Upsilon(2S,3S)\to\gamma\cpoddhiggs$,
$\cpoddhiggs\to\mu^+\mu^-$. Such a particle
appears in extensions of the Standard
Model, where a light \CP-odd Higgs boson naturally couples strongly to
$b$-quarks. 
We find no evidence for such processes in the mass range 
$0.212\le m_{A^0}\le9.3$\,GeV
in the samples of $99\times10^6$ \Y2S\ and $122\times10^6$ \Y3S\
decays collected by the \babar\ detector at the \pep2\
B-factory and set
stringent upper limits on the effective coupling of the $b$ quark to
the \cpoddhiggs. 
We also limit the dimuon branching fraction of the $\eta_b$
meson: $\BR(\eta_b\to\mu^+\mu^-)<0.9\%$ at 90\% confidence level.
\end{abstract}

% PACS, the Physics and Astronomy Classification Scheme.

\maketitle

% The body of the paper starts here
%%%%%%%%%%%%%%%%%%%%%%%%%%%%%%
% INTRODUCTION
%%%%%%%%%%%%%%%%%%%%%%%%%%%%%%

The concept of mass is one of the most intuitive ideas in physics
since it is present in everyday human experience.
Yet the fundamental nature of mass 
remains one of the great mysteries of science.  The Higgs mechanism is a
theoretically appealing way to account for the different masses of elementary 
particles~\cite{ref:Higgs}. It 
implies the existence of at least one new scalar particle, 
the Higgs boson, which is the only Standard Model
(SM)~\cite{ref:SM} particle yet to be observed. 
%% Its discovery would have a profound effect  
%% on our fundamental understanding of matter.
The SM Higgs boson mass is constrained to be of
$\mathcal{O}(100-200\,\mathrm{GeV})$ by  
direct searches~\cite{Barate:2003sz}
%\cite{Herndon:ICHEP08},
and by
precision electroweak measurements~\cite{LEPSLC:2005ema}.

%% The Standard Model and the simplest electroweak symmetry breaking
%% scenario suffer from quadratic divergences in the radiative
%% corrections to the mass parameter of the Higgs potential. 
%% Several theories beyond the Standard Model that regulate these
%% divergences have been proposed. 
%% Supersymmetry~\cite{ref:SUSY} is one such model; however, in its simplest form 
%% (the Minimal Supersymmetric Standard Model, MSSM) questions of
%% parameter fine-tuning and ``naturalness'' of the Higgs mass scale
%% remain. 

A number of theoretical models extend the Higgs sector to 
include additional Higgs fields, some of
them naturally light~\cite{Dermisek:2005ar}.
Similar light scalar states, \eg\ axions, appear in models motivated
by astrophysical 
observations and are typically assumed to have Higgs-like
couplings~\cite{NomuraThaler}. Direct searches typically constrain 
the mass of such a light particle, \cpoddhiggs, to be below
$2m_b$~\cite{Dermisek:2006}, making it 
accessible to radiative decays of $\Upsilon$
resonances~\cite{Wilczek:1977zn}. Model predictions for the branching
fraction (BF) of $\Upsilon\to\gamma\cpoddhiggs$ decays range from
$10^{-6}$~\cite{Dermisek:2006py,NomuraThaler} to as high as
$10^{-4}$~\cite{Dermisek:2006py}. 
Empirical
motivation for a low-mass Higgs search comes from the HyperCP
experiment~\cite{HyperCP}, which observed three anomalous events
in the $\Sigma^+\to p\mu^+\mu^-$ final state. These events have been 
interpreted as production of a scalar boson with the mass of
$214.3$\,\mev decaying 
into a pair of muons~\cite{XJG}.
The large datasets available at \babar\ allow us to place stringent
constraints on such models. 

If a light scalar \cpoddhiggs\ exists, the pattern of its decays would depend
on its mass. 
Assuming no invisible (neutralino) decays~\cite{BAD2073},
for low masses $\higgsmass<2 m_\tau$ the BF $\BR_{\mu\mu}\equiv\BR(\cpoddhiggs\to\mu^+\mu^-)$,
should be sizable. 
Significantly above the $\tau$ threshold,
$\cpoddhiggs\to\tau^+\tau^-$ would dominate, and 
hadronic decays might also 
be significant. 

This Letter describes a search for a resonance in the
dimuon invariant mass distribution for the fully reconstructed final state
$\Upsilon(2S,3S)\to\gamma\cpoddhiggs,\ \cpoddhiggs\to\mu^+\mu^-$. We
assume that the decay width of 
the resonance is negligibly 
small compared with the experimental resolution, as
expected~\cite{NomuraThaler,ref:Lozano} for \higgsmass\ sufficiently
far from the mass of the $\eta_b$~\cite{ref:etab}. 
We further assume that the resonance is a scalar (or pseudo-scalar)
particle. While the significance of any observation would not depend on this
assumption, the signal efficiency and, therefore, the BFs
are computed for a spin-0 particle. 
In addition, following
the recent discovery of the $\eta_b$ meson~\cite{ref:etab}, we look
for the leptonic decay of the $\eta_b$ 
through $\Upsilon(2S,3S)\to\gamma\eta_b$,
$\eta_b\to\mu^+\mu^-$. 
%% If the recently discovered state is the
%% conventional quark-antiquark $\eta_b$ meson, its leptonic width is
%% expected to be negligible. Thus, setting a limit on the dimuon
%% branching fraction sheds some light on the nature of the recently
%% discovered state. 
We use $\Gamma(\eta_b)=10\pm5$\,MeV, the range
expected in most theoretical models and  
consistent with the \babar\ results~\cite{ref:etab}. 

%%%%%%%%%%%%%%%%%%%%%%%%%%%%%%
%THE \babar\ DETECTOR AND DATASET
%%%%%%%%%%%%%%%%%%%%%%%%%%%%%%
%
We search for
two-body transitions $\Upsilon(2S,3S)\to\gamma\cpoddhiggs$, followed by 
decay $\cpoddhiggs\to\mu^+\mu^-$ 
 in samples of $(98.6\pm0.9)\times10^6$ \Y2S and 
$(121.8\pm 1.2)\times10^6$ \Y3S\
decays collected with the \babar\ detector
at the \pep2\ asymmetric-energy \epem\ collider at the SLAC National
Accelerator Laboratory. 
We use a sample of $79\,\mathrm{fb}^{-1}$
accumulated on the \Y4S resonance (\Y4S sample) for studies of the continuum
backgrounds. Since the \Y4S is three orders of magnitude broader than
the \Y2S and \Y3S, the BF $\BR(\Y4S\to\gamma\cpoddhiggs)$ is expected to
be negligible. 
For characterization of the background events and selection
optimization, we also use a sample of
$1.4\,\mathrm{fb}^{-1}$ ($2.4\,\mathrm{fb}^{-1}$) collected $30$\,MeV
below the \Y2S (\Y3S) resonance (off-resonance samples).
The \babar\ detector is described in detail
elsewhere~\cite{detector,LST}. 

%%%%%%%%%%%%%%%%%%%%%%%%%%%%%%
% EVENT SELECTION
%%%%%%%%%%%%%%%%%%%%%%%%%%%%%%

We select events with exactly two oppositely-charged tracks and a
single energetic photon with a center-of-mass (CM) energy
$E^{*}_\gamma\ge0.2$\,\gev, while allowing additional photons 
with CM energies below $0.2$\,\gev to be present in the  event. We assign a
muon mass hypothesis to the two tracks (henceforth referred to as muon
candidates), and require that at least one is positively identified
as a muon~\cite{LST}.
We require
that the muon candidates form a geometric vertex with $\chi^2_\mathrm{vtx}<20$
for 1 degree of freedom and
displaced transversely by 
at most $2$\,cm~\cite{dvtxCut} from the nominal location of the
$e^+e^-$ interaction 
region. We perform 
a kinematic fit to the $\Upsilon$ candidate formed from the two muon
candidates and the energetic photon. The CM energy of the
$\Upsilon$ candidate is constrained, within the beam energy spread, to
the total 
beam energy $\sqrt{s}$, and the decay vertex of the $\Upsilon$ is
constrained to the beam interaction region. 
We select events with $-0.2<\sqrt{s}-m(\Upsilon)<0.6$\,\gev and 
place a requirement on the kinematic fit $\chi^2_{\Upsilon}<30$ (for 6
degrees of freedom). 
We further  
require that the momenta of the dimuon candidate
$\cpoddhiggs$ and the photon are back-to-back in the CM frame to
within $0.07$ rad, and that the cosine of the angle
between the muon direction and $\cpoddhiggs$ direction in the center
of mass of the $\cpoddhiggs$ is less than $0.92$.  
The selection criteria are chosen to maximize $\varepsilon/\sqrt{B}$,
where $\varepsilon$ is the average selection efficiency for a broad
\higgsmass range and $B$ is the background yield
in the off-resonance sample.  

\begin{figure}[h!]
\begin{center}
\subfigure[]{\epsfig{file=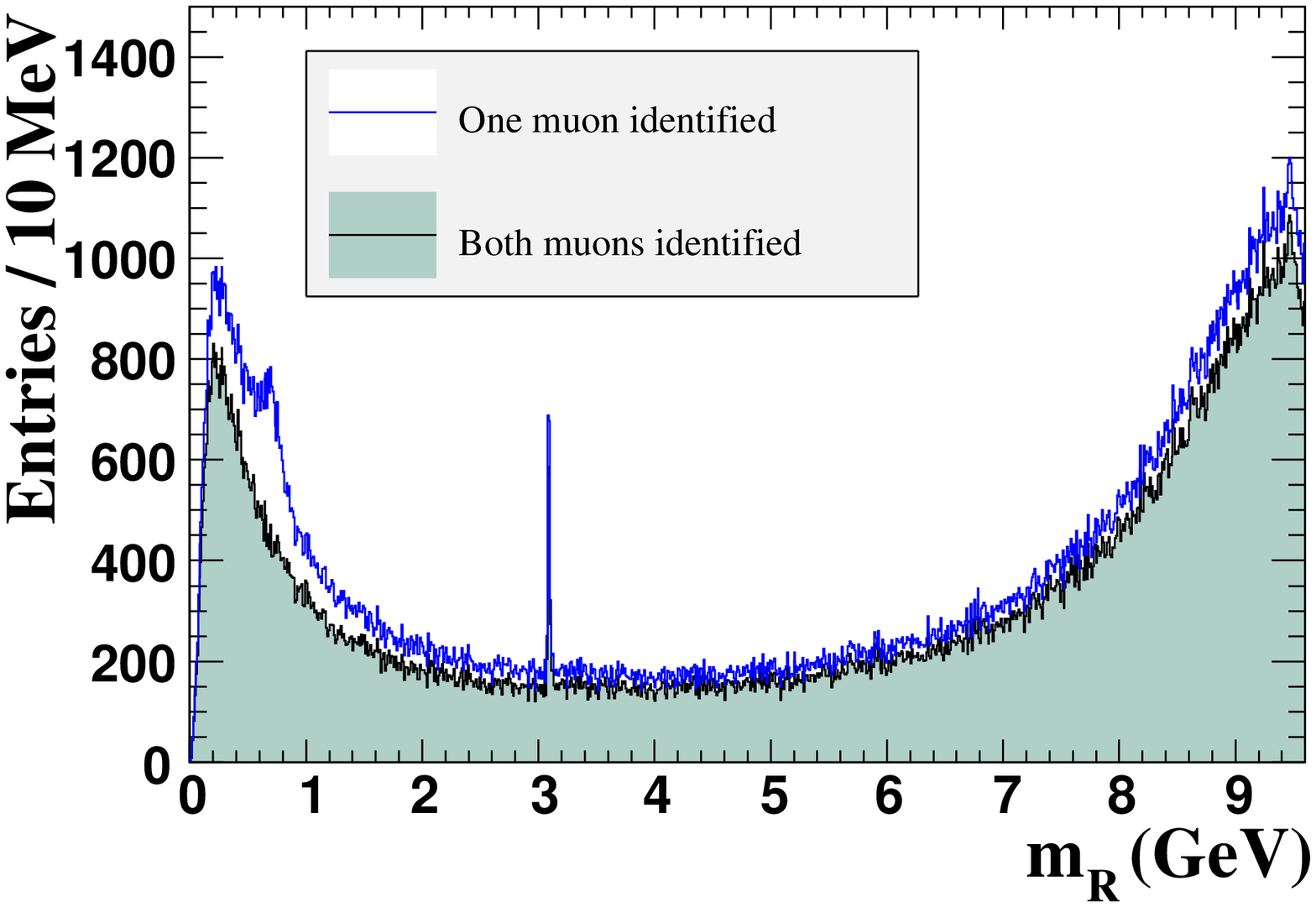,width=0.45\textwidth}}\hspace{0.1 in}
\subfigure[]{\epsfig{file=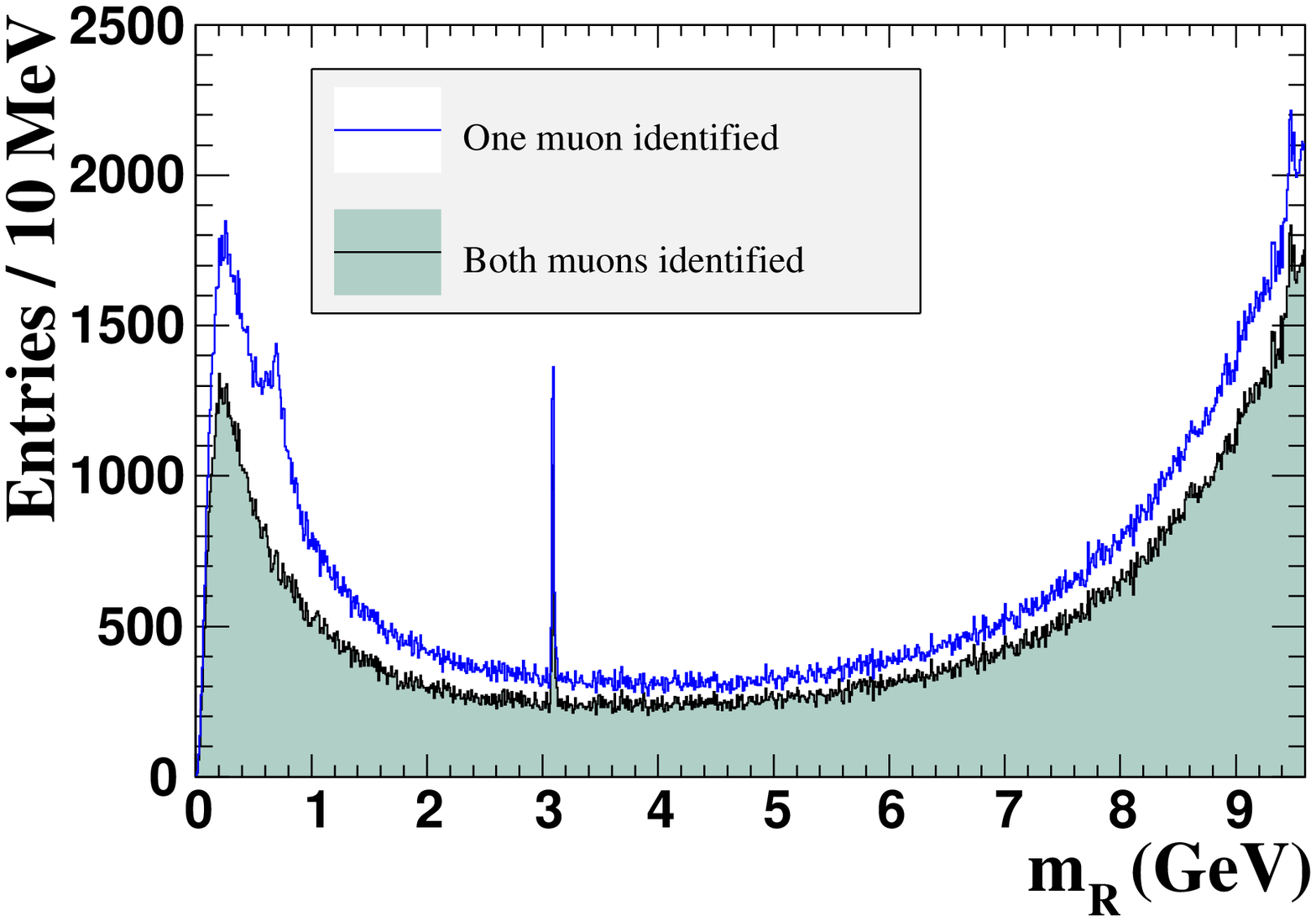,width=0.45\textwidth}}
\end{center}
\caption{Distribution of the reduced mass \redmass\ in
  (a) the \Y2S data and (b) the \Y3S data. Blue (open) histogram shows the
  distribution for the 
  selection in which only one of two muons is required to be
  positively identified. The peak from
  $e^+e^-\to\gamma_\mathrm{ISR}\rho^0(770)$, $\rho^0\to\pi^+\pi^-$, in
  which one of the pions is misidentified as a muon, is clearly
  visible. 
  Black (filled) histogram shows the distribution for the
  selection in which both muons are
  positively identified (this selection is used in the search for
  $0.5\le\higgsmass<1.05$~GeV). The ISR-produced peaks
  $J/\psi\to\mu^+\mu^-$ and $\Y1S\to\mu^+\mu^-$ are visible for both
  selections.  
}
\label{fig:spectrum}
\end{figure}

The criteria above select
387,546 \Y2S and 724,551
\Y3S events 
(mass spectra for \Y2S and \Y3S datasets are shown in 
Fig.~\ref{fig:spectrum} in \cite{EPAPS}). 
The backgrounds are dominated by two 
types of QED processes: ``continuum'' $e^+e^-\to\gamma\mu^+\mu^-$
and the initial-state radiation (ISR) production of 
$\rho^0$, $\phi$, $J/\psi$, $\psi(2S)$, and \Y1S vector mesons. In order to suppress
contributions from the 
ISR-produced 
$\rho^0\to\pi^+\pi^-$ final state in which a
pion is misidentified as a muon (probability $\sim$3\%/pion), we require that
both tracks are positively identified as muons when we search for \cpoddhiggs
candidates in the range
$0.5\le\higgsmass<1.05$\,\gev. 
Finally, when selecting candidate events in
the $\eta_b$ region with dimuon invariant mass 
$m_{\mu\mu}\sim 9.39$\,\gev in the \Y2S 
(\Y3S) dataset, we 
suppress the decay chain
$\Y2S\to\gamma_2\chi_b(1P),\ \chi_b(1P)\to\gamma_1\Y1S$ 
($\Y3S\to\gamma_2\chi_b(2P),\ \chi_b(2P)\to\gamma_1\Y1S$)
by requiring that
no secondary photon $\gamma_2$ above a CM energy of $E_2^*=0.1$\,\gev
($0.08$\,\gev) is present in the event. 

We use signal Monte Carlo (MC) samples~\cite{geant,EvtGen}
$\Y2S\to\gamma\cpoddhiggs$ and 
$\Y3S\to\gamma\cpoddhiggs$ generated at 20 values of \higgsmass over 
a broad range $0.212\le\higgsmass\le9.5$\,GeV to
measure the selection efficiency for the signal 
events. The efficiency varies between 24--55\%, depending \higgsmass.

%%%%%%%%%%%%%%%%%%%%%%%%%%%%%%
% EXTRACTION OF SIGNAL YIELDS
%%%%%%%%%%%%%%%%%%%%%%%%%%%%%%

We extract the yield of signal events as a function of 
\higgsmass\ in the interval $0.212\le\higgsmass\le 9.3$\,GeV by
performing a series of unbinned extended maximum likelihood fits
to the distribution of the reduced mass
$\redmass\equiv\sqrt{m_{\mu\mu}^2 - 4m_\mu^2}$.
The likelihood function contains contributions from signal, continuum
background, and, where appropriate, peaking backgrounds, as described
below. 
For $0.212\le\higgsmass<0.5$\,GeV, 
we fit over a fixed interval $0.01<\redmass<0.55$\,GeV; near the \jpsi
resonance, we fit over the interval $2.7<\redmass<3.5$\,GeV; and near the 
$\psi(2S)$ resonance we fit over the range $3.35<\redmass<4.1$\,GeV. 
Elsewhere, we use sliding intervals
$\mu-0.2<\redmass<\mu+0.1$\,GeV, where $\mu$ is the mean of the signal
distribution of $\redmass$. We search for \cpoddhiggs\ in fine
mass steps $\Delta\higgsmass=2$--$5$\,MeV. We sample a total of 1951
\higgsmass values. For each \higgsmass\ value, we determine the BF
products
$\BR_{nS}\equiv\BR(\Upsilon(nS)\to\gamma\cpoddhiggs)\times\BR_{\mu\mu}$,
where $n=2,3$. 
Both the fitting procedure and the event
selection were developed and tested using MC and \Y4S samples prior
to their application to the \Y2S\ and \Y3S\ data sets. 

The signal probability density function (PDF) is described by a sum of
two Crystal Ball 
functions~\cite{ref:CBshape} 
with tail parameters on either side of the maximum. The
signal PDFs are centered around the expected values of
\redmass and have a typical
resolution of 2--10\,MeV, which increases monotonically with \higgsmass. 
We determine the PDF as a function of
$\higgsmass$ using the signal MC samples, and we
interpolate PDF parameters and signal efficiency values 
linearly between the simulated
points. We determine the uncertainty in the PDF parameters by
comparing the distributions of the simulated and reconstructed 
$e^+e^-\to\gamma_\mathrm{ISR}J/\psi,\ J/\psi\to\mu^+\mu^-$
events.

We describe the continuum background below $\redmass<0.23$\,\gev with 
a threshold function 
$f_\mathrm{bkg}(\redmass)\propto\tanh\left(\sum_{\ell=1}^3p_\ell\redmass^\ell\right)$. 
The parameters 
$p_\ell$ are fixed to the values determined from the fits to the 
$e^+e^-\to\gamma\mu^+\mu^-$ MC sample~\cite{KK2F} and
agree, within statistics, with those determined by fitting 
the \Y2S, \Y3S, and \Y4S samples with the signal contribution set to zero. 
Elsewhere the
background is well described in each limited \redmass range by a first-order
($\redmass<9.3$\,GeV) or a second-order ($\redmass>9.3$\,GeV)
polynomial with coefficients determined by the fit. 

Events due to known resonances $\phi$, \jpsi, $\psi(2S)$, and \Y1S are
present in our 
sample in specific \redmass intervals, and constitute peaking
  backgrounds.  
We include these contributions in the fit where appropriate, and describe the
shape of the resonances using the same functional form as for the signal,
a sum of two 
Crystal Ball functions, with parameters determined from fits to
the combined \Y2S and \Y3S dataset. The contribution to the event
yield from 
$\phi\to K^+K^-$, in which one of the kaons is misidentified as a
muon, is fixed to $111\pm24$ (\Y2S) and $198\pm42$ (\Y3S). We
determine this contribution from the event yield 
of $e^+e^-\to\gamma\phi,\ \phi\to K^+K^-$ in a sample where both kaons are positively
identified, corrected for the measured misidentification rate of kaons as
muons. 
We do not search for \cpoddhiggs candidates in the immediate
vicinity of \jpsi and $\psi(2S)$, excluding regions of $\pm40$\,MeV
around \jpsi ($\approx\pm5\sigma$) and $\pm25$\,MeV
($\approx\pm3\sigma$) around $\psi(2S)$. 

%%%%%%%%%%%%%%%%%%%%%%%%%%%%%%%%%%%%%%%%
% SYSTEMATIC UNCERTAINTIES
%%%%%%%%%%%%%%%%%%%%%%%%%%%%%%%%%%%%%%%%

We compare the overall
selection efficiency between the data and the MC simulation by
measuring the absolute cross section $d\sigma/d\redmass$ for the
radiative QED process
$e^+e^-\to\gamma\mu^+\mu^-$ over the broad kinematic range
$0<\redmass\le9.6$\,GeV, using the off-resonance sample. We use the
ratio of measured to expected~\cite{KK2F} cross 
sections to correct the signal selection efficiency as a function of
\higgsmass. This correction ranges between 4--10\%, with a systematic
uncertainty of 5\%. 
This uncertainty accounts for effects of selection, 
reconstruction (for both charged tracks and the photon),
and trigger efficiencies. 

We determine the uncertainty in the signal and peaking background PDFs
by comparing the distributions of $\approx 4000$ data 
and MC $e^+e^-\to\gamma_\mathrm{ISR}J/\psi,\,J/\psi\to\mu^+\mu^-$
events. We correct for the observed difference in the width of
the \redmass distribution ($5.3$\,MeV in MC 
versus $6.6$\,MeV in the data)
and use half of the correction to estimate 
the systematic uncertainty on the signal yield. This is the dominant
systematic uncertainty 
on the signal yield for $\higgsmass>0.4$\,\gev. 
We estimate that the uncertainties in the tail parameters of the
Crystal Ball PDF contribute less than 1\% to the uncertainty in signal
yield based on the observed variations in the $J/\psi$ yield. 
The systematic uncertainties due to the fixed continuum background PDF for 
$\redmass<0.23$ and the fixed contribution from
$e^+e^-\to\gamma\phi$ do not
exceed $\sigma_\mathrm{bkg}(\BR_{nS})=0.2\times10^{-6}$. These are the largest
systematic contributions for $0.212\le\higgsmass<0.4$~GeV. 

We test for possible bias in the fitted value of the signal yield with a
large ensemble of pseudo-experiments. 
The bias is
consistent with zero for all values of \higgsmass, and we assign a
BF uncertainty of $\sigma_\mathrm{bias}(\BR_{nS})=0.05\times10^{-6}$
at all values of \higgsmass to cover the
statistical variations in the results of the test. 

%%%%%%%%%%%%%%%%%%%%%%%%%%%%%%
% STATISTICAL INTERPRETATION %
%%%%%%%%%%%%%%%%%%%%%%%%%%%%%%

\begin{figure}[tb]
\begin{center}
\epsfig{file=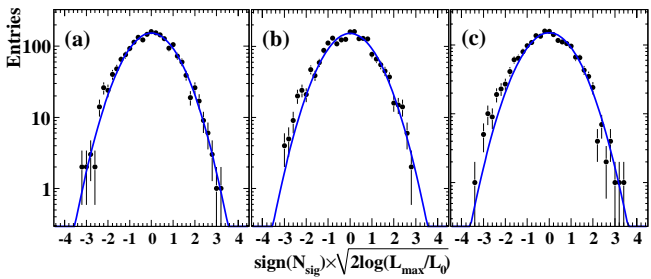, width=0.48\textwidth}
\end{center}
\vspace{-2\baselineskip}
\caption{Distribution of the log-likelihood variable
  $\mathcal{S}$ with both statistical and systematic uncertainties
  included for (a) \Y2S\ fit, (b) \Y3S\ fit, and (c) combination of
  \Y2S\ and \Y3S data. There are no points outside of displayed
  region of $\mathcal{S}$. 
  The solid curve is the standard normal distribution.
}
\label{fig:scan1d_sign_Run7}
\end{figure}

To estimate the significance of any positive fluctuation, 
we compute the likelihood ratio variable 
$\mathcal{S}(\higgsmass) = \mathrm{sign}(N_\mathrm{sig})\sqrt{2\log(L_{\max}/L_0)}$,
where $L_{\max}$ is the maximum
likelihood value for a fit with a free signal yield centered at
\higgsmass, $N_\mathrm{sig}$ is that fitted signal yield, and 
$L_0$ is the value of the likelihood for the signal yield fixed at
zero. 
Under the null hypothesis
$\mathcal{S}$ is expected to be normal-distributed with $\mu=0$ and
$\sigma=1$
(Fig.~\ref{fig:scan1d_sign_Run7}). 
Including systematics, the largest $\mathcal{S}$ values are $3.1$ (\Y2S)
and $2.8$ (\Y3S), consistent with
a null-hypothesis distribution for 1951 \higgsmass points. 
%

%%%%%%%%%%%%%%%%%%%%%%%%%%%%%%
% RESULTS AND CONCLUSIONS
%%%%%%%%%%%%%%%%%%%%%%%%%%%%%%

\begin{figure}[tb]
\begin{center}
\epsfig{file=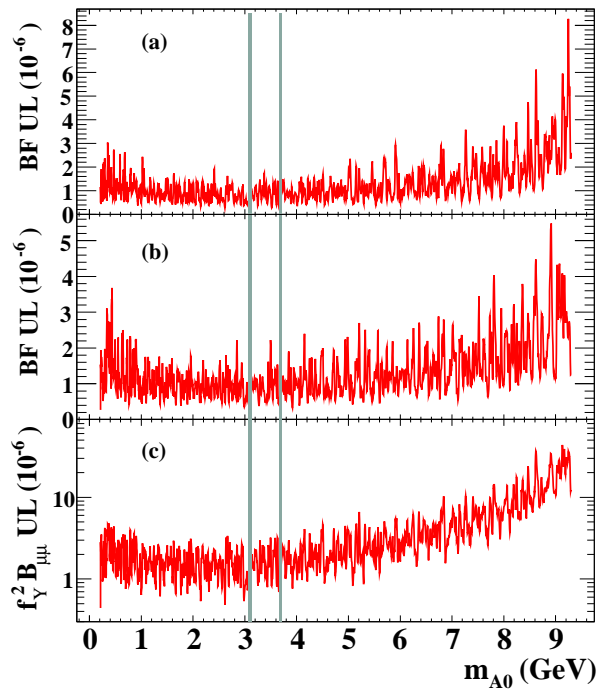, width=0.48\textwidth}
\end{center}
\vspace{-2\baselineskip}
\caption{90\% C.L. upper limits on (a) $\BR(\Y2S\to\gamma A^0)\times\BR_{\mu\mu}$, 
  (b) $\BR(\Y3S\to\gamma A^0)\times\BR_{\mu\mu}$, 
  and (c) effective coupling $f_{\Upsilon}^2\times\BR_{\mu\mu}$
  as a function of \higgsmass. 
  The shaded areas show the regions
  around the \jpsi and $\psi(2S)$ resonances excluded from
  the search. 
}
\label{fig:scan1d_limits}
\end{figure}

Since we do not observe a significant excess of events above the background
in the range $0.212<\higgsmass\le 9.3$\,GeV, we set 
upper limits on $\BR_{2S}$ and $\BR_{3S}$. 
We add statistical and systematic uncertainties 
in quadrature. 
The 90\% confidence level (C.L.) Bayesian upper limits,
computed with a uniform prior and assuming a Gaussian likelihood
function, are shown in Fig.~\ref{fig:scan1d_limits} as a function of mass
$\higgsmass$. The limits vary 
from $0.26\times10^{-6}$ to $8.3\times10^{-6}$ ($\BR_{2S}$) and 
from $0.27\times10^{-6}$ to $5.5\times10^{-6}$ ($\BR_{3S}$). 

\begin{figure}[h!]
\begin{center}
\epsfig{file=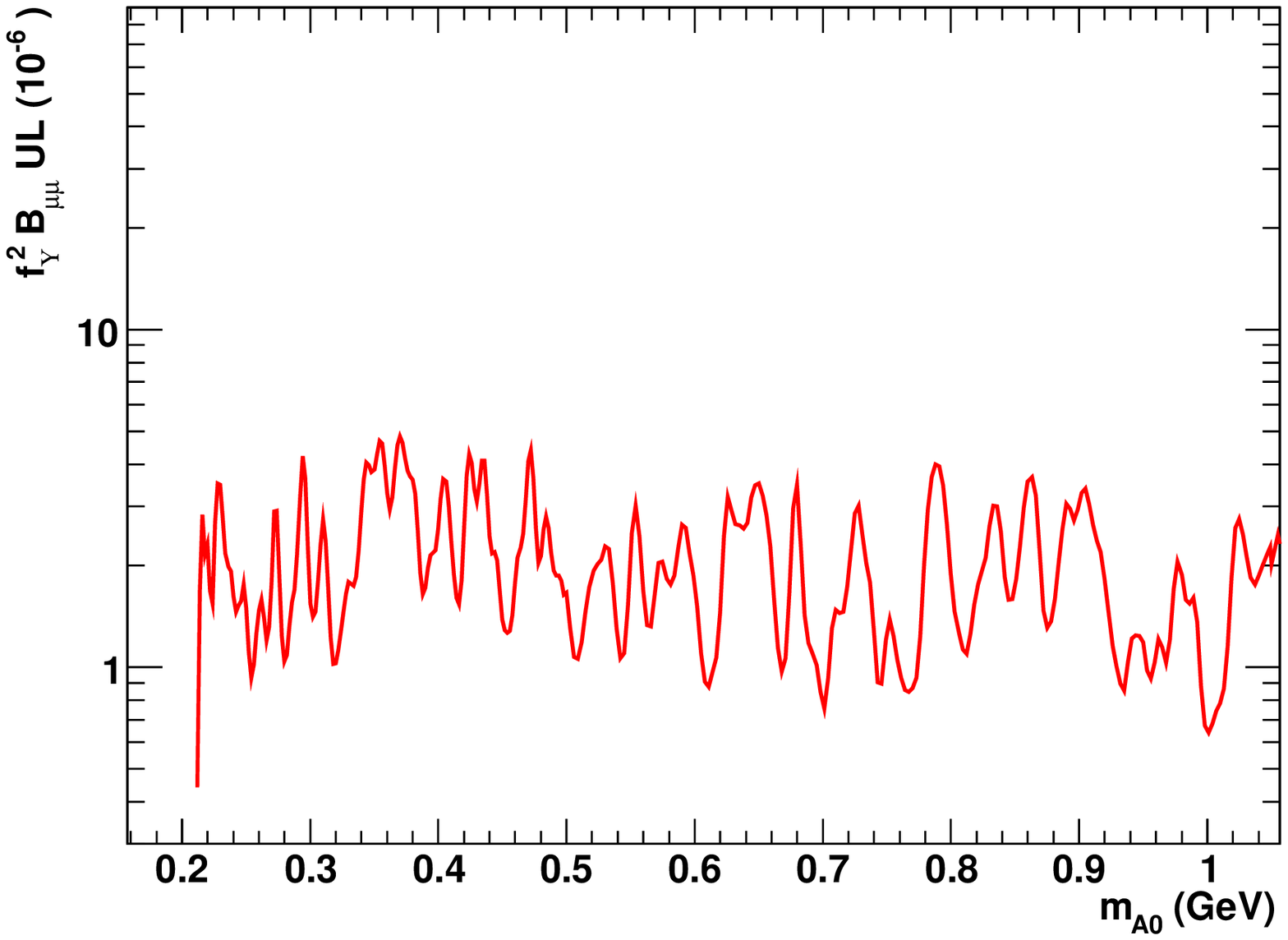, height=3in}
\end{center}
\caption{90\% C.L. upper limits on the effective Yukawa coupling
  $f^2_\Upsilon\times\BR_{\mu\mu}$ as a function of \higgsmass  in the range
  $0.212\le\higgsmass\le1.05$~\gev.
}
\label{fig:scan1d_comb_yukawa_low}
\end{figure}

\begin{figure}[h!]
\begin{center}
\epsfig{file=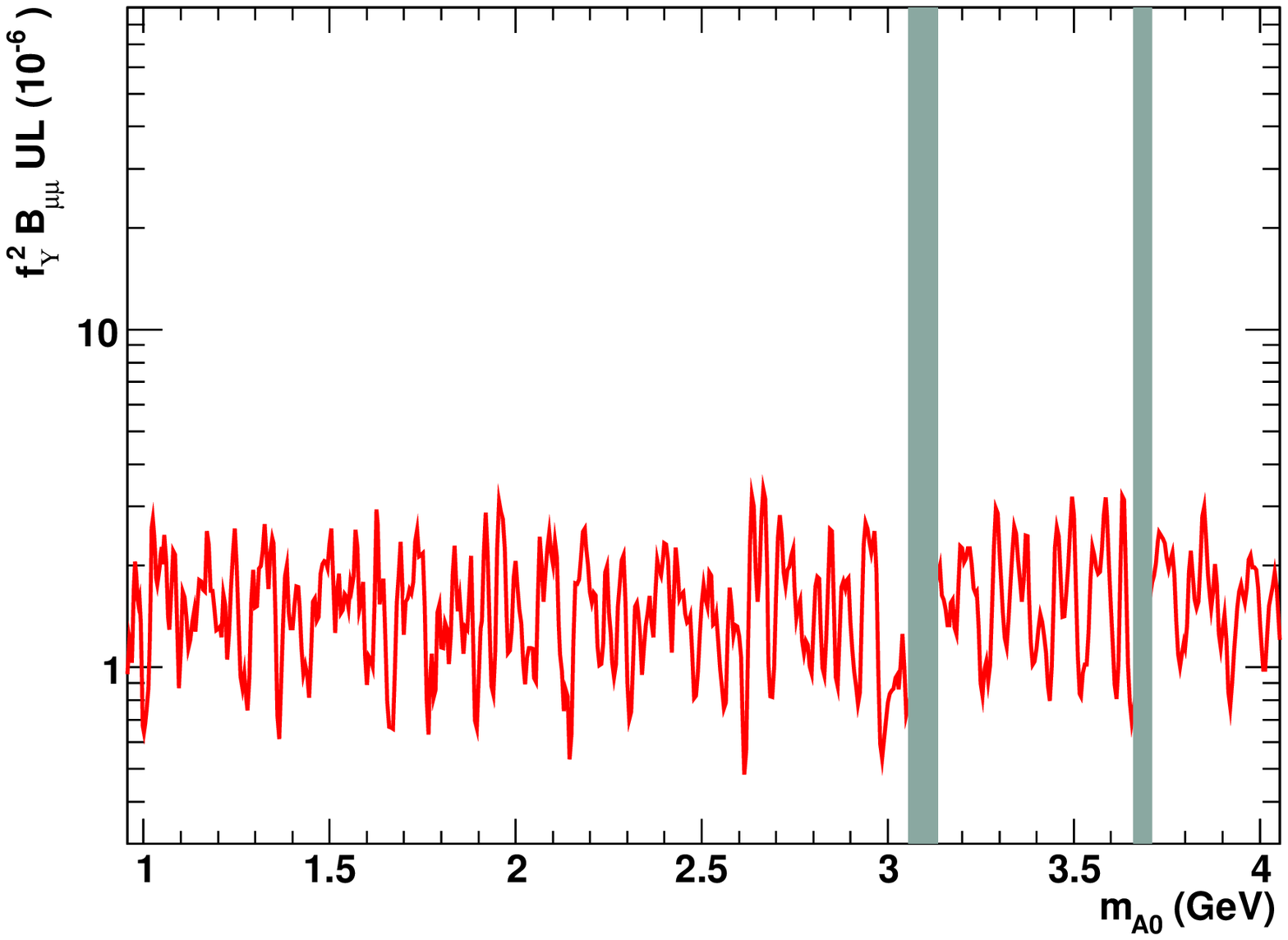, height=3in}
\end{center}
\caption{Upper limits on the effective Yukawa coupling
  $f^2_\Upsilon\times\BR_{\mu\mu}$ as a function of \higgsmass   in the range
  $1\le\higgsmass\le4$~\gev.
  The shaded areas show the regions
  around the \jpsi and $\psi(2S)$ resonances excluded from
  the search. 
}
\label{fig:scan1d_comb_yukawa_med}
\end{figure}

\begin{figure}[h!]
\begin{center}
\epsfig{file=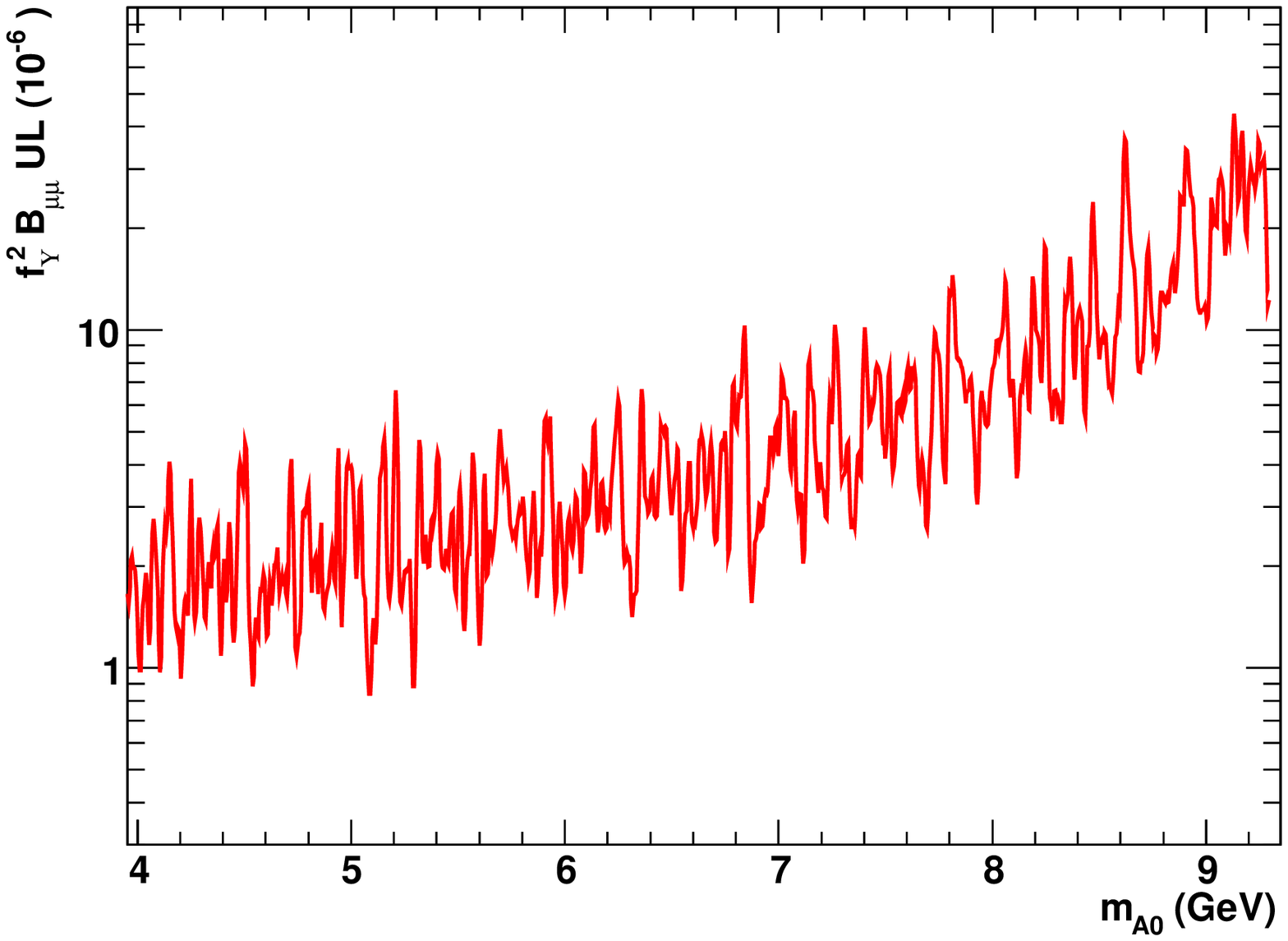, height=3in}
\end{center}
\caption{Upper limits on the effective Yukawa coupling
  $f^2_\Upsilon\times\BR_{\mu\mu}$  as a function of \higgsmass  in the range
  $4\le\higgsmass\le 9.3$~\gev.
}
\label{fig:scan1d_comb_yukawa_high}
\end{figure}

The BFs $\BR(\Upsilon(nS)\to\gamma\cpoddhiggs)$ are
related to the effective 
coupling $f_{\Upsilon}$ of the bound $b$-quark to the $A^0$
through~\cite{Wilczek:1977zn,Nason86}: 
\beq
\frac{\BR(\Upsilon(nS)\to\gamma\cpoddhiggs)}{\BR(\Upsilon(nS)\to l^+l^-)} = 
\frac{f_{\Upsilon}^2}{2\pi\alpha}\left(1-\frac{m^2_{\cpoddhiggs}}{m^2_{\Upsilon(nS)}}\right)
\label{eq:BFRatio}
\eeq
where $l\equiv e$ or $\mu$ and $\alpha$ is a fine structure constant.
%computed at the scale $m_{\Upsilon(nS)}$. 
The effective coupling $f_{\Upsilon}$ includes 
the Yukawa coupling of the $b$-quark and 
the \higgsmass-dependent QCD and
relativistic corrections to $\BR_{nS}$~\cite{Nason86} and 
the leptonic width of $\Upsilon(nS)$~\cite{Barbieri75}. To first order
in $\alpha_S$, the corrections range from 0 to 30\%~\cite{Nason86} but have
comparable uncertainties~\cite{Beneke1997}. The ratio of corrections
for \Y2S and \Y3S is within 4\% of unity~\cite{Nason86} in the
relevant range of \higgsmass. We do not attempt to
factorize these contributions, but instead compute the
experimentally-accessible quantity $f_{\Upsilon}^2\BR_{\mu\mu}$ and
average \Y2S\ and \Y3S results, taking into account both correlated
and uncorrelated uncertainties. 
The combined upper limits are shown as a function of \higgsmass in
Fig.~\ref{fig:scan1d_limits}(c) 
(plots with expanded mass scales in three ranges of
\higgsmass  are available in
Fig.~\ref{fig:scan1d_comb_yukawa_low}-\ref{fig:scan1d_comb_yukawa_high}
in \cite{EPAPS})
and span the range 
$(0.44-44)\times10^{-6}$, at 90\% C.L. The combined likelihood
variable 
$\langle\mathcal{S}\rangle=(w_{2S}\mathcal{S}_{2S}+
w_{3S}\mathcal{S}_{3S})/\sqrt{w_{2S}^2+w_{3S}^2}$
is shown in Fig.~\ref{fig:scan1d_sign_Run7}c, where $w_{nS}$ is the
statistical weight of the $\Upsilon(nS)$ dataset in the average. The
largest fluctuation 
is $\langle\mathcal{S}\rangle=3.3$. 
Our set of 1951 overlapping fit regions corresponds to $\approx 1500$
independent measurements~\cite{ref:CONF}.  We determine the
probability to observe a 
fluctuation of $\langle\mathcal{S}\rangle=3.3$ or larger in such a
sample to be at least 45\%.

\begin{figure}[tb]
\begin{center}
\subfigure[]{\epsfig{file=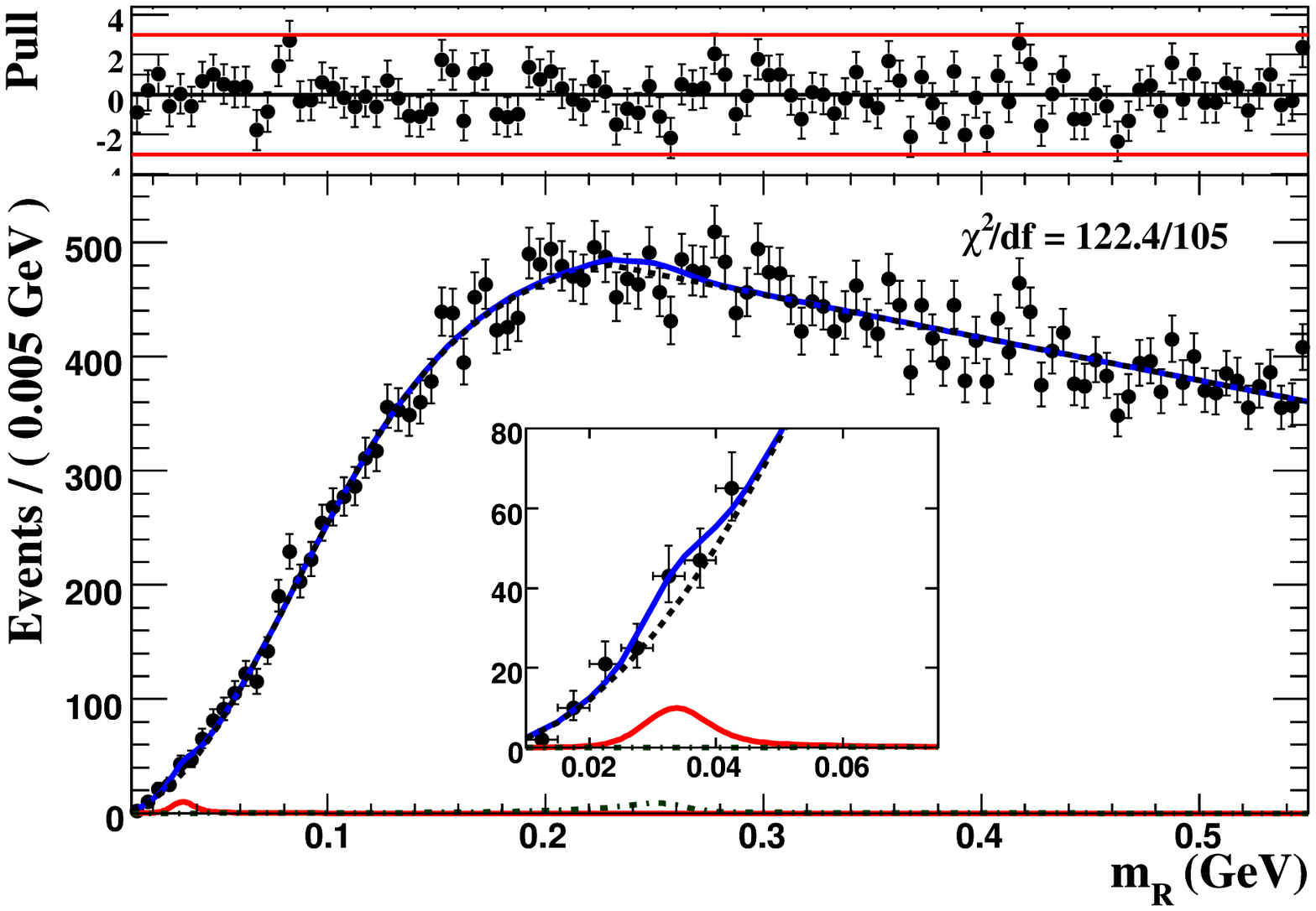,width=0.45\textwidth}}\hspace{0.1 in}
\subfigure[]{\epsfig{file=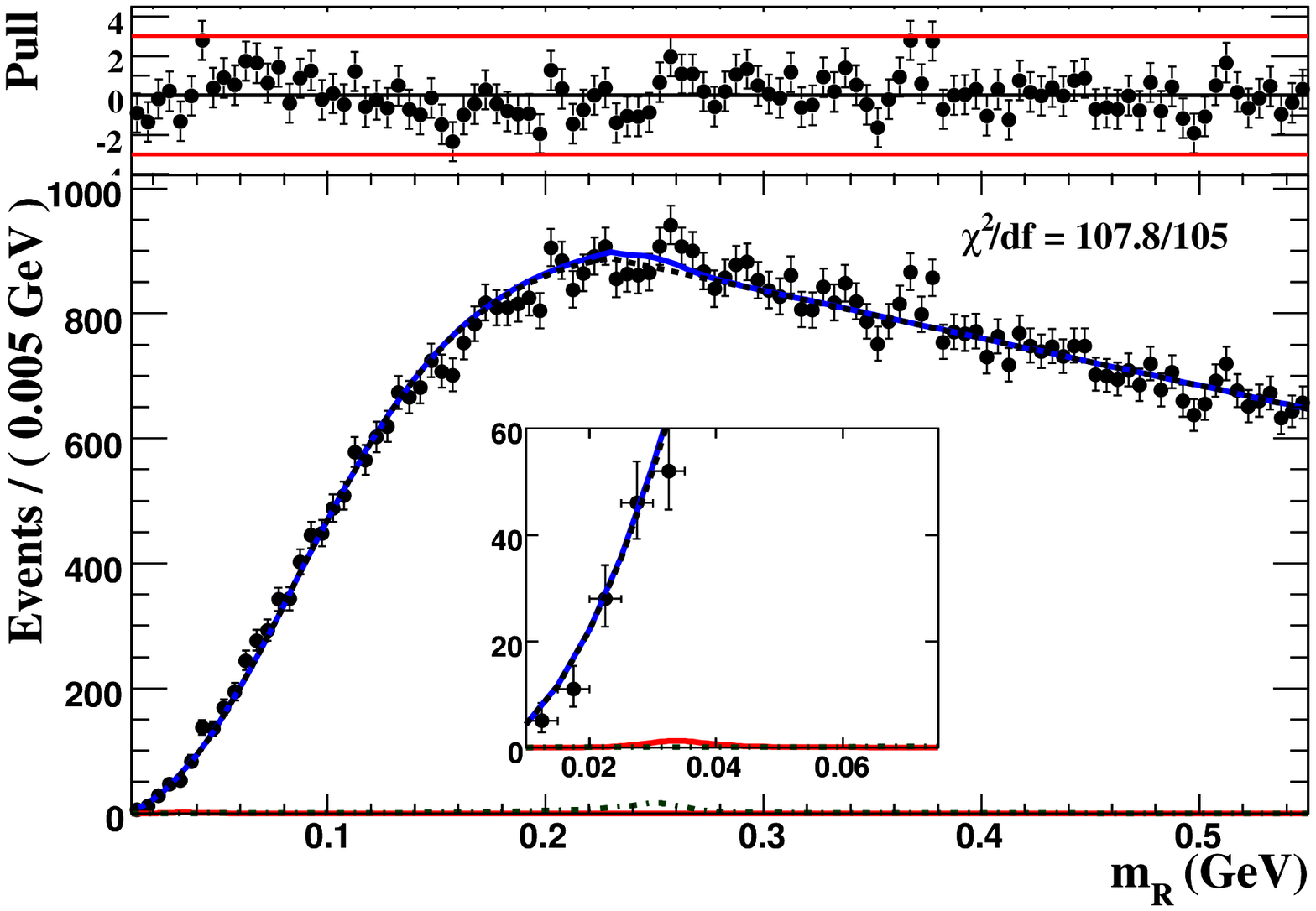,width=0.45\textwidth}}
\end{center}
\caption{Fits to the HyperCP region $\higgsmass=0.214$~\gev in
  (a) \Y2S\ dataset and (b) \Y3S\ dataset. The bottom graph shows the
  $\redmass$ distribution 
  (solid points), overlaid by the full fit (solid blue
  line). Also shown are the contributions from the signal at
  $\higgsmass=0.214$~\gev (solid red line) and the continuum
  background (dashed black line). The inset zooms in on the signal
  region. 
  The top plot shows the normalized residuals
  $p=(\mathrm{data}-\mathrm{fit})/\sigma(\mathrm{data})$  with unit
  error bars. The individual fits correspond to the log-likelihood
  ratios of  $\mathcal{S}=2.0$ (\Y2S) and  $\mathcal{S}=0.2$ (\Y3S),
  and the combined significance $\langle\mathcal{S}\rangle=1.4$. 
}
\label{fig:proj1d_0.214}
\end{figure}

We do not observe any significant signal at
$\higgsmass=0.214$\,\gev (Fig.~\ref{fig:proj1d_0.214} in \cite{EPAPS})
and set an upper limit on the coupling  
$f^2_{\Upsilon}(\higgsmass=0.214\,\mathrm{GeV})<1.6\times10^{-6}$ at
90\% C.L (assuming $\BR_{\mu\mu}=1$), which is significantly smaller
than the value required to explain the HyperCP events as light
Higgs production~\cite{XJG}. 

\begin{figure}[tb]
\begin{center}
\subfigure[]{\epsfig{file=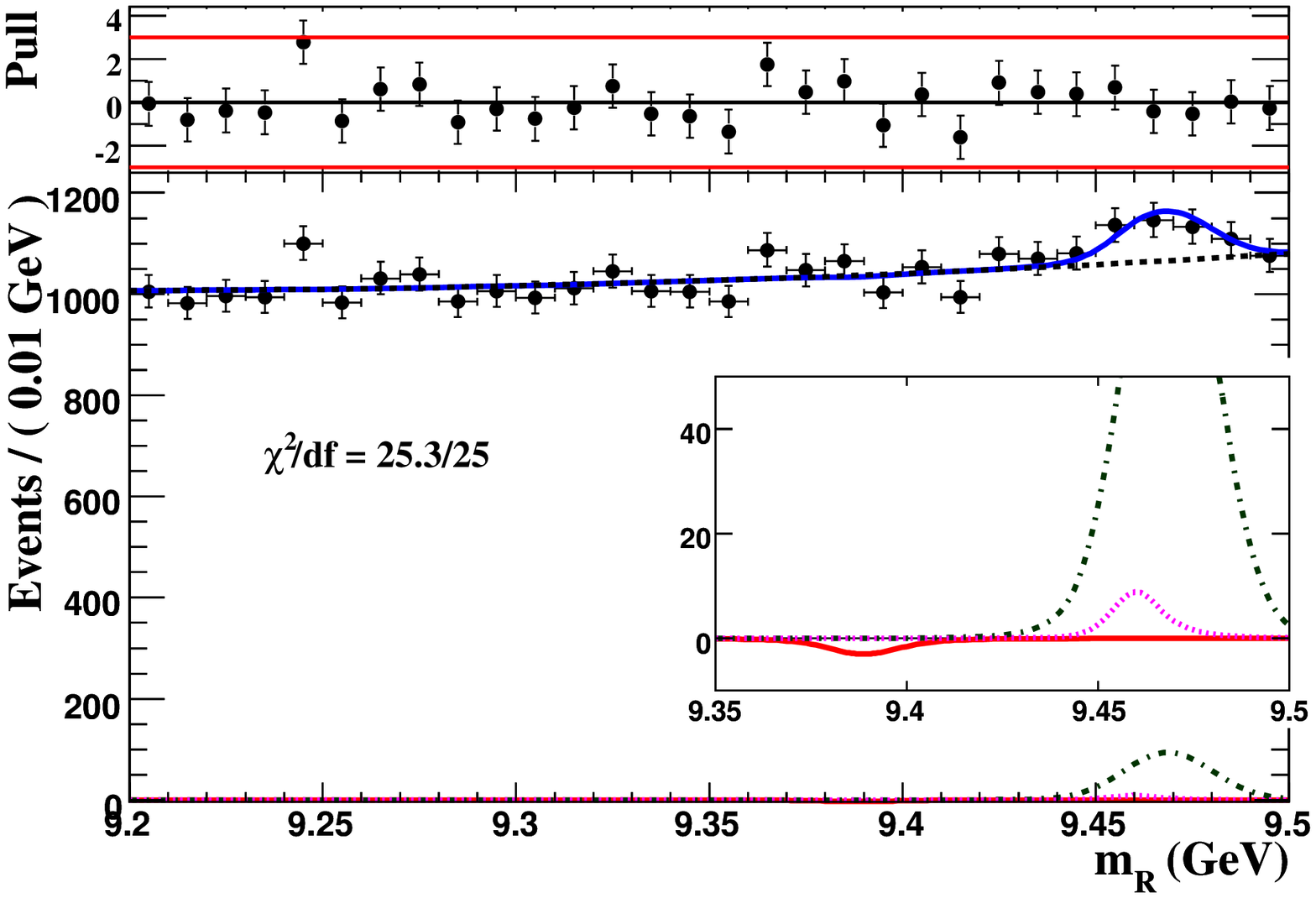,width=0.45\textwidth}}\hspace{0.1 in}
\subfigure[]{\epsfig{file=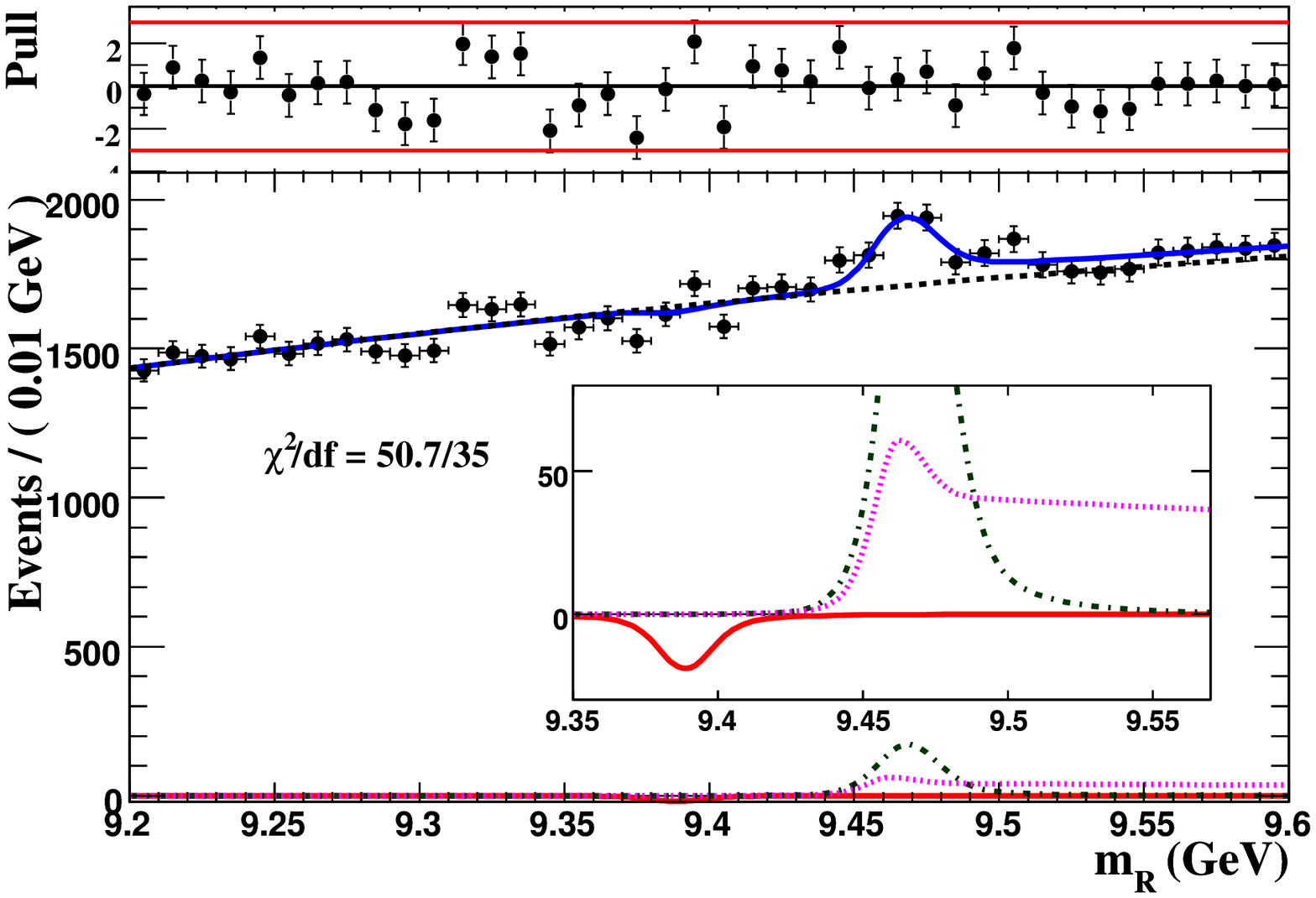,width=0.45\textwidth}}
\end{center}
\caption{Fits to the $\eta_b$ region in (a)
  \Y2S\ dataset and (b) \Y3S\ dataset. The bottom graph shows the
  $\redmass$ distribution 
  (solid points), overlaid by the full fit (solid blue
  line). Also shown are the contributions from the signal at
  at $m_{\eta_b}=9.389$~\gev (solid red line), background from the
  $e^+e^-\to\gamma_\mathrm{ISR}\Y1S$ (dot-dashed green line),
  background from $\Y3S\to\gamma\chi_{b}(2P)$,
  $\chi_{b}(2P)\to\gamma\Y1S$ (dotted magenta line), and the continuum
  background (dashed black line). The inset zooms in on the signal
  region. 
  The top plot shows the normalized residuals
  $p=(\mathrm{data}-\mathrm{fit})/\sigma(\mathrm{data})$  with unit
  error bars. 
}
\label{fig:etab_run7}
\end{figure}

A fit to the $\eta_b$ region (Fig.~\ref{fig:etab_run7} in
\cite{EPAPS}) includes background contributions from
the ISR process $e^+e^-\to\gamma_\mathrm{ISR}\Y1S$, and from the cascade
decays $\Upsilon(nS)\to\gamma_2\chi_{bJ},\,\chi_{bJ}\to\gamma_1\Y1S$ 
with $\Y1S\to\mu^+\mu^-$. We measure the rate of the ISR events in the
\Y4S dataset, scale it to the \Y2S and \Y3S data, and fix this contribution
in the fit. The rate of the cascade decays, the number of signal
events, and the continuum background are free in the fits to
the \Y2S and \Y3S data sets. 
We measure 
$\BR(\Y2S\to\gamma\eta_b)\times\BR(\eta_b\to\mu^+\mu^-)=(-0.4\pm3.9\pm1.4)\times10^{-6}$
and 
$\BR(\Y3S\to\gamma\eta_b)\times\BR(\eta_b\to\mu^+\mu^-)=(-1.5\pm2.9\pm1.6)\times10^{-6}$,
where the first uncertainty is statistical and the second is
systematic, dominated by the uncertainty in $\Gamma(\eta_b)$. 
Taking into account the \babar\
measurements of $\BR(\Y2S\to\gamma\eta_b)$
and $\BR(\Y3S\to\gamma\eta_b)$~\cite{ref:etab}, we derive
$\BR(\eta_b\to\mu^+\mu^-)=(-0.25\pm0.51\pm0.33)\%$ and 
$\BR(\eta_b\to\mu^+\mu^-)<0.9\%$ at 90\% C.L. This limit is consistent
with the mesonic interpretation of the $\eta_b$
state. 

In summary, we find no evidence for the dimuon decays of a light
scalar particle 
in radiative decays of \Y2S\ and \Y3S\ mesons. 
We set upper
limits on the coupling $f^2_{\Upsilon}\times\BR_{\mu\mu}$ for
$0.212\le\higgsmass\le 9.3$~GeV. Assuming $\BR_{\mu\mu}\approx 1$ in
the mass range  
$2m_\mu\le\higgsmass\le1$~GeV, our results limit the coupling
$f_{\Upsilon}$ to be at most 12\% of the Standard Model coupling of
the $b$ quark to the Higgs boson. 
Our limits rule out much of the
parameter space allowed by the light Higgs~\cite{Dermisek:2006py} and
axion~\cite{NomuraThaler} models. We also set an upper limit on the
dimuon branching fraction of the $\eta_b$. 

%

%%%%%%%%%%%%%%%%%%%%%%%%%%%%%%
% ACKNOWLEDGMENTS
%%%%%%%%%%%%%%%%%%%%%%%%%%%%%%

% Standard acknowledgments paragraph; must always be included.
We are grateful for the excellent luminosity and machine conditions
provided by our \pep2\ colleagues, 
and for the substantial dedicated effort from
the computing organizations that support \babar.
The collaborating institutions wish to thank 
SLAC for its support and kind hospitality. 
This work is supported by
DOE
and NSF (USA),
NSERC (Canada),
CEA and
CNRS-IN2P3
(France),
BMBF and DFG
(Germany),
INFN (Italy),
FOM (The Netherlands),
NFR (Norway),
MES (Russia),
MEC (Spain), and
STFC (United Kingdom). 
Individuals have received support from the
Marie Curie EIF (European Union) and
the A.~P.~Sloan Foundation.

\onecolumngrid
\newpage

\section{Appendix: EPAPS Material}

The following includes supplementary material for the Electronic
Physics Auxiliary Publication Service. 

%%\subsection{Additional Plots}

%


\begin{thebibliography}{99}
\bibitem{EPAPS}
Additional plots are available in the Appendix.
\end{thebibliography}
\end{document}